\documentclass[conference]{IEEEtran}
\usepackage{url}
\usepackage{graphicx}
\usepackage{caption}

\usepackage{cite}

\ifCLASSINFOpdf
\else
\fi
\begin{document}
%
\title{A proposal of a methodological framework with experimental guidelines to investigate clustering stability on financial time series}

\author{\IEEEauthorblockN{Gautier Marti, Philippe Very \\  Philippe Donnat}
\IEEEauthorblockA{Hellebore Capital Management \\
63 Avenue des Champs-\'Elys\'ees \\
firstname.lastname@helleborecapital.com}
\and
\IEEEauthorblockN{Frank Nielsen}
\IEEEauthorblockA{Laboratoire d'Informatique de l'\'Ecole polytechnique\\
Palaiseau, FRANCE\\
nielsen@lix.polytechnique.fr}}


%


\maketitle

\begin{abstract}
We present in this paper an empirical framework motivated by the practitioner point of view on stability. The goal is to both assess clustering validity and yield market insights by providing through the data perturbations we propose a multi-view of the assets' clustering behaviour.
The perturbation framework is illustrated on an extensive credit default swap time series database available online at \url{www.datagrapple.com}. 
\end{abstract}


%
\IEEEpeerreviewmaketitle

\section{Introduction}
Clustering, the task of grouping a set of objects in such a way that objects in the same group, also called cluster, are more similar to each other than those in different groups, is not yet a common tool for financial time series analysis. However, its application area seems wide: portfolio diversification \cite{tola2008cluster} and risk analysis, investment strategies such as statistical arbitrage. This unsupervised machine learning technique suffers from several drawbacks which must be dealt with before being adopted by the financial community. These algorithms act as black boxes whose properties such as consistency, i.e. the probability that a given algorithm outputs the target clustering converges to 1, and rate of convergence are generally unknown in the asymptotic regimes which matter in finance, i.e. Random Matrix Theory Asymptotics \cite{laloux2000random}, High Dimension Moderate Sample Size Asymptotics (HDMSS) and High Dimension Low Sample Size Asymptotics (HDLSS) \cite{borysov2014asymptotics}.
Besides lacking from these theoretical properties for now, clusterings obtained are much dissimilar from one algorithm to another when applied on financial time series \cite{lemieux2014clustering, musmeci2014relation}. Even worse, adding small noise to a given sample and applying the same algorithm on the original one and its perturbated version yields different clusters. These current shortcomings prevent robust cluster-based applications in finance where data is noisy observations of the underlying phenomenon. For instance, a trading strategy based on clustering events such as emergence, combination, split and decay \cite{sun2007graphscope,tantipathananandh2011finding} of clusters is flawed if these events spuriously occur due to the algorithm unstability.

Concerning the object of study, namely Credit Default Swaps (CDS), they are vanilla credit derivatives (cf. \cite{hull2006options} for an introduction to derivatives and \cite{o2011modelling} for a thorough treatment of credit derivatives and CDS mathematical modelling) whose purpose is to transfer credit risk from a counterparty to another. They may be considered as an insurance against the default (or other credit events defined precisely by an ISDA contract) of the underlying entity, usually a corporate, a financial or a sovereign entity. The price of this insurance can thus be viewed as a fear gauge expressing the market apprehension of an entity default risk: higher the risk, higher the price to pay to get insured. The CDS market (being an over-the-counter (OTC) market) was accused to lack transparency, today it is still difficult to gather public information on this market. We can mention DTCC which reports traded volumes once a week, and DataGrapple which displays historical price time series of 700 entities from 2006 to present updated on a daily basis.

\subsection{Related Work}

Clustering financial time series was mainly explored in econophysics \cite{tumminello2005tool, tola2008cluster, tumminello2010correlation} where it is considered as a competitive way to the Random Matrix Theory (RMT) \cite{laloux2000random,bun2015rotational} to filter the correlation matrix.
It was noticed (for instance, in \cite{musmeci2014relation}) that clustering algorithms recover roughly industrial classification benchmarks (ICB).
Authors propound to benchmark these algorithms against an ICB, and measure the similarity between a resulting clustering and the chosen ICB using the Adjusted Rand Index (ARI). They consider this measure as a proxy of the information filtered by the methods, and thus consider it as a way to perform model selection by opting for the clustering method with the highest score.
Yet, a statistical analysis such as clustering can produce clusters which may not exist in an industrial classification: for example, the PIIGS cluster (assets from Portugal, Ireland, Italy, Greece and Spain) during the European debt crisis. In \cite{musmeci2014relation}, authors also use bootstrapping \cite{efron1979bootstrap} to assess the robustness of their clustering algorithm using the intuitive but moot idea of clustering stability \cite{carlsson2010characterization}, i.e. the reproducibility of the clustering when data is slightly perturbed. Clustering stability for model selection or validity assessment \cite{levine2001resampling} is indeed a hot topic in the machine learning literature: \cite{ben2006sober} warn against its irrelevant use as stability only depends on the uniqueness of the clustering objective function minimizer for large sample size, yet \cite{shamir2007cluster} advocate that this criterion remains useful in the case of finite possibly small samples.
From the quantitative analyst or trader perspective, dynamical stability of models, i.e. stability to online perturbations arising from streaming data, is required for being confident using them. A clustering should only change when a meaningful event happens in the market. Despite being an important notion for practitioners, few works have dealt with the dynamical stability properties of clusters computed on financial time series \cite{di2010use}.
Clustering CDS has not been widely explored in the literature since data is scarce and difficult to obtain. \cite{mayordomo2010all} investigate the differences in the main data sources employed by researchers and policymakers and conclude that credit default swap databases are not all equal. Some of these databases are built using hypothetical prices resulting from an average of data collected but on which, possibly, no trade could have been done. 
We will use the database from DataGrapple which is built tick by tick from many market dealers. This allows to make a synthetic order book of best bid / best offer across the dealers' prices at a given time $t$ improving the confidence that a trade could have effectively been done at this level at time $t$.

\subsection{Contributions}

We propose an empirical framework to investigate clustering on financial time series and obtain insights from the perturbations which are motivated by financial applications. To illustrate the proposed perturbations, we apply them to study a large CDS dataset of market prices. We compare on this dataset the clusterings obtained using several distance matrices $D_{ij}$:
\begin{itemize}
\item $D_{ij}^{P} = (1-\rho_{ij})/2$, where $\rho_{ij}$ is the Pearson correlation coefficient,
\item $D_{ij}^{S} = (1-\rho_{S_{ij}})/2$, where $\rho_{S_{ij}}$ is the Spearman correlation,
\item $D_{ij}^{E}$ is an Euclidean distance,
\item $D_{ij}^{G}$ is a distance designed for working on i.i.d. random processes  \cite{marti2015toward}.
\end{itemize}  
Our experimental study advocates that:
\begin{itemize}
\item if dependence between assets is the only focus of study, $D_{ij}^{S}$ (Spearman) should be preferred to $D_{ij}^{P}$ (Pearson); $D_{ij}^{E}$ should not be used for comparing financial time series; finally, $D_{ij}^{G}$ is the most versatile distance and behaves well under all proposed perturbations, it mostly behaves like $D_{ij}^{S}$ and should be preferred;
\item clustering can be persistent through a wide range of perturbations provided that an appropriate distance is leveraged by the clustering algorithm.
\end{itemize}
Besides, we observe some interesting facts about the CDS market and illustrate clustering stability by leveraging the Sankey diagram from the Data-Driven Documents library.

\section{Perturbation Framework for Financial Time Series}
\label{perturb}

Perturbations can be performed both on prices (some of the $T$ values of each time series) or on assets (some of the $N$ time series themselves). Concretely, these perturbations consist in modifying row-wise or column-wise the $N \times T$ data matrix $X$.

\subsection{Time Perturbation}

We provide below a list of perturbations concerning some of the $T$ time series values. We explain their motivations arising from financial concerns and what we can learn from them by analyzing the clustering stability.

\subsubsection*{Sliding Window}

\underline{Motivation:} Dynamical stability of models is a requirement for trading and risk information systems. For example, value at risk (VaR), an estimated amount of money so that the potential loss of a portfolio over a given timespan should not exceed, is computed on a moving window and updated with respect to the asset prices stream. With no trading in the portfolio, and in a stationary regime, VaR should not vary too much.

\underline{Definition:} Given a window width $W$ and a step size $S$, clustering is performed on $X_{:,[t_{\mathrm{cur}},t_{\mathrm{cur}}+W[}$ and $X_{:,[t_{\mathrm{cur}}+S,t_{\mathrm{cur}}+S+W[}$, then current time $t_{\mathrm{cur}}$ is updated $t_{\mathrm{cur}} := t_{\mathrm{cur}} + S$, and so on.

\underline{Insight:} A clustering that strongly differs from one time to another when the market seems in a steady regime should be rejected since very sensitive to noise, i.e. small unsignificant market variations. If confidence is high in the methodology, modification of clusters may be a signal that market structure is changing (for instance, end of a crisis and decrease in global correlation).

\subsubsection*{Odd vs. Even}

\underline{Motivation:} A clustering algorithm applied on two samples describing the same phenomenon should yield the same results. How to obtain two of these samples? The goal is to split the sample in two while mitigating the effect of non-stationarity, seasonality, end-of-the-week trading activity, meetings and announcements from the ECB or the FED generally happening on Friday.

\underline{Definition:} We define $X^{(1)} = \{X_{:,t} ~|~ t\mathrm{~is~odd}\}$ and $X^{(2)} = \{X_{:,t} ~|~ t\mathrm{~is~even}\}$, i.e. we build the sample of the odd trading days and the sample of the even trading days. Since the trading week lasts 5 days, we alleviate the aforementioned statistical biases. Clustering is performed independently on $X^{(1)}$ and on $X^{(2)}$.

\underline{Insight:} If not stable, the clustering method should be rejected.

\subsubsection*{Economic Regimes}

\underline{Motivation:} Since the economic context can change dramatically, financial time series do not evolve in a steady regime in the long run. It makes sense to split the timespan into different periods where the statistical regime can be considered stationary.

\underline{Definition:} We partition the sample into $M$ subsamples $X = \sqcup_{i=1}^{M} X^{(i)}$, where $X^{(i)} = X_{:,[t_i,t_{i+1}[}$ and the $(t_i)_{i=1}^{M+1}$ delimit time intervals. The breakpoints $(t_i)_{i=1}^{M+1}$ can be chosen guided by market understanding or computed using a dynamical changes and regime detection algorithm.

\underline{Insight:} We can study whether the clustering structure is persistent throughout different economic regimes, and how strongly it is. If not, which are the periods concerned and how steep is the change? Which assets are involved? Why do they switch from clusters? What happened to them and their clusters? However, we must keep in mind that it is difficult to separate the signal from the noise of the clustering methodology.

\subsubsection*{Heart vs. Tails}

\underline{Motivation:} Does the market under stress share a common clustering structure with the market during uneventful periods?

\underline{Definition:} Let $\overline{X}_t = \frac{1}{N} \sum_{i=1}^N X_{i,t}$ be the mean time series of the market. let $Q_1$ be the lower quartile and $Q_3$ be the upper quartile. We define $\mathcal{T} = \{t ~|~ \overline{X}_t \leq Q_1 \lor \overline{X}_t \geq Q_3\}$ and $\mathcal{H} = \{t ~|~ \overline{X}_t \notin \mathcal{T}\}$ corresponding to times having market values in the tails and in the heart respectively. Then we split the sample $X$ in the two following subsamples 
$X^{(1)} = \{X_{:,t} ~|~ t \in \mathcal{T}\}$ and $X^{(2)} = \{X_{:,t} ~|~ t \in \mathcal{H}\}$ on which we apply the same clustering algorithm.

\begin{figure}
\centering
\includegraphics[scale=0.48]{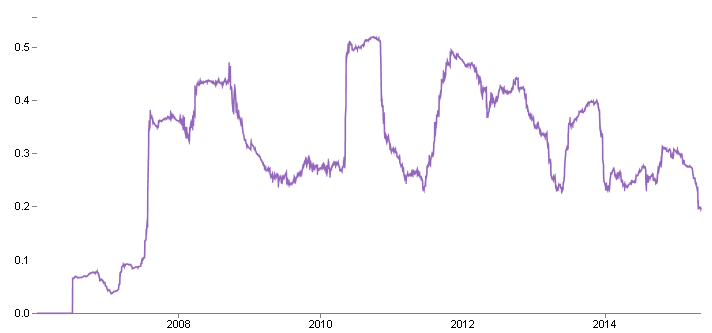}
\caption{Mean Correlation Dynamics computed on the whole CDS dataset from 2006 to 2015 using a 6-month sliding window}\label{fig:mean_correl}
\end{figure}

\underline{Insight:} Although it is difficult to anticipate changes of the market behaviour, in period of stress all assets tend to be simultaneously affected by macroeconomic tensions which usually induces a significant increase in correlation betweem them (cf. Fig.~\ref{fig:mean_correl}). Thus, correlation should be less discriminating and a correlation-based clustering might be unstable with respect to this perturbation.

\subsubsection*{Multiscale}

\underline{Motivation:} Markets prices can be monitored from a high frequency sampling (tick by tick or minute by minute) to much lower frequency (from hours to hours, days by days or on a weekly basis). The sampling frequency used is linked to the type of trading, from high-frequency trading (HFT) and algorithmic trading to long-term investments. Is the clustering structure persistent throughout a wide range of time scales or does it strongly depend on the sampling?

\underline{Definition:} From $X$ we can build $M$ datasets $X^{(i)} = X_{:,s_i}$, $i = 1,\ldots,M$ where $s_i$ are regularly-spaced multiscale subsample of $\{1,\ldots,T\}$.

\underline{Insight:} Ideally, in the perspective of building a risk system, we would like that the choice of risk factors is independent of the time scale used for the analysis. This perturbation allows us to verify to which extent clustering is multiscale persistent.

%
%
%
%

\subsubsection*{Maturities}

\underline{Motivation:} Fixed-income assets such as bonds, swaps and CDS for instance, have a lifespan called their maturity. Several products with different maturities (say an insurance against the default of a corporate for 1, 2, 3, 5, 7 or 10 years) can concern the same entity. Since the underlying risk is the same, we would like similar clusterings.

\underline{Definition:} We get from the market several time series dataset $X^{(i)}$, one $N \times T$ data matrix for each quoted maturity.

\underline{Insight:} We can either reject a clustering method which yields to unstable clusters or investigate why a particular maturity has a different clustering structure compared to the others.

\subsubsection*{Term Structure}

\underline{Motivation:} The term structure is the set of all quoted maturities. Clustering term structure could lead to a more meaningful result than clustering separately each maturity.

\underline{Definition:} For clustering term structures, one need a specific distance. To our knowledge, the problem of obtaining a proper one which captures the whole information (e.g. dynamics, distribution and correlation of its distortions) has not been addressed. Here, we give a simple one for clustering CDS term structures at a given date $t$. A CDS probability of default $P(t)$ can be viewed as a cumulative distribution function on $\mathbf{R}^+$. Indeed, the probability of default is increasing, the probability of instantaneous default is 0, and at infinity all entities will eventually default. Thus, $f(t) = \partial P(t)/\partial t$ defines a probability density function on $\mathbf{R}^+$, and since $\int_{\mathbf{R}^+} f(t) ~dt = 1$, $\sqrt{f(t)}$ is a unit vector in $L^2(\mathbf{R})$. The inner product between two unit vectors defines an angle $\phi$ which is the distance between two term structures. Given two term structures $P_1$, $P_2$ and $f_1$, $f_2$ such that $f_i(t) = \partial P_i(t)/\partial t$, their distance $\phi$ can be written $\cos \phi = \int_{\mathbf{R}^+} \sqrt{f_1(t)} \sqrt{f_2(t)} ~dt = 1 - H^2(\sqrt{f_1(t)},\sqrt{f_2(t)})$, where $H$ is the Hellinger distance.

\underline{Insight:} For entities near default, the term structure should be inverted, i.e. the market anticipates a renormalisation if these entities survive. For entities having seemingly no troubles, the quoted term structure should mirror the debt term structure of these entities. Some industries have a particular debt structure (short term debt for financials, long term debt for basic materials and industrials). Part of this information should also be captured by correlation between assets on a given maturity.


%
%
%

\subsection{Population Perturbation}

To the presented time-based perturbations, we add the following two population-based perturbations on the set of assets: \textbf{increasing/decreasing the number of entities} and \textbf{adding entities with imputed historical prices}. These perturbations can be easily motivated: new companies emerge regularly and some others disappear from the market. The clustering structure should not radically change when adding or removing entities from the clustering perimeter. When new companies are created and introduced in the market, they have not much history. It may be necessary to impute missing data based, for instance, on a clustering methodology. We would like to verify that adding synthetic time series built from existing ones to the clustering perimeters does not change the original clustering structure. The clustering structure should be robust to the statistical engineering performed to impute missing data or clean their poor quality.

\section{Insights from Benchmarking on Stability}
\label{bench}

\subsection{CDS Dataset}

We apply the proposed perturbation framework on a dataset of Credit Default Swap (CDS) time series which can be seen at \url{www.datagrapple.com}. This dataset have 450 entities which have complete historical daily prices from January 2006 to today, i.e. more than 2300 daily prices for a given asset (say Nokia Oyj) on a given maturity (say a 5 year contract), and 250 more with missing data (for instance, Numericable Group S.A.) whose missing historical prices where imputed using a machine learning algorithm. Since credit default swaps are traded over-the-counter, closing time for fixing prices can be arbitrarily chosen, here 5pm GMT, i.e. after the London Stock Exchange trading session. This synchronous fixing of CDS prices avoids spurious correlations arising from different closing times. For example, the use of close-to-close stock prices artificially overestimates intra-market correlation and underestimates inter-market dependence since they have different trading hours \cite{martens2001returns}. 
Moreover, besides being publicly available, another benefits of this dataset is that historical prices are mid-market data computed from a synthetic real-time order book of best bid / best offer that are proposed by the main CDS market makers instead of averaged consensus prices on which it could have been impossible to trade.

\subsection{Preprocessing of Financial Time Series}

Given $N$ time series of prices $P_i(t)$, $i = 1,\ldots,N$, $t = 1,\ldots,T$, the standard approach consists in working with the time series of log-returns $\Delta \log P_i(t) = \log P_i(t+1) - \log P_i(t)$. This transform mitigates the risk of spurious correlation by empirically stationarizing the considered time series besides normalizing the variations. Thus, clustering the $N$ time series $\Delta \log P_i(t)$, $t = 1,\ldots,T$ using a distance based on Pearson correlation or an Euclidean distance may be more relevant.
In \cite{marti2015toward}, authors claim that the standard Euclidean distance should not be used for clustering financial time series. For instance, assuming that variations of two time series $X$ and $Y$ are independent and identically distributed according to $\mathcal{N}(\mu_X,\sigma_X^2)$ and $\mathcal{N}(\mu_Y,\sigma_Y^2)$, we have $\mathbf{E}[(X-Y)^2] = (\mu_X - \mu_Y)^2 + (\sigma_X - \sigma_Y)^2 + 2\sigma_X \sigma_Y(1 - \rho(X,Y))$. If $X$ and $Y$ are independent, then $\rho(X,Y) = 0$, and $\mathbf{E}[(X-Y)^2] = (\mu_X - \mu_Y)^2 + \sigma_X^2 + \sigma_Y^2$. Now, suppose that $\mu_X = \mu_Y$ and $\sigma_X = \sigma_Y$, the distance $2\sigma_X^2$ between $X$ and $Y$ grows proportionally to the variance which is not a desirable behaviour for two time series whose variations have the same distribution. Based on the observation that the Euclidean distance mixes correlation and distribution information but inappropriately for their purpose, authors have designed a distance of the form $D_{\mathrm{correlation}} +  D_{\mathrm{distribution}}$ for comparing i.i.d. random processes both on correlation and distribution information. Using their distance we can work directly on $\Delta P_i(t) = P_i(t+1) - P_i(t)$ since the log-transform is useless: both the dependency (based on Spearman correlation) and distribution (Hellinger) distance employed are invariant to monotonic transforms of the data. Thus, we will work on $\Delta \log P_i(t)$ when clustering using $D^P$ and $D^E$, and on $\Delta P_i(t)$ when clustering using $D^S$ and $D^G$. In the remaining, we apply for both experiments a hierarchical clustering, namely weighted linkage, with a constant height cut to obtain $K = 16$ clusters; results presented still holds for various values of $K$ and different clustering algorithms.

\subsection{Comparison of Distances using the Perturbation Framework}

In this section, we leverage the proposed perturbation framework to test four distances used for clustering financial time series. We also observe some stylized facts about the CDS market. The four distances $D^P$, $D^S$, $D^E$ and $D^G$ are essentially distances based on Pearson correlation, Spearman correlation, Euclidean distance and a recently introduced distance based on correlation and distribution respectively. We illustrate the clustering stability with a data visualization, a Sankey diagram, which highlights the dissimilarties between partitions \cite{marti2015hcmapper}.

\begin{figure}
   \begin{minipage}[c]{.24\linewidth}
      \includegraphics[width=\textwidth]{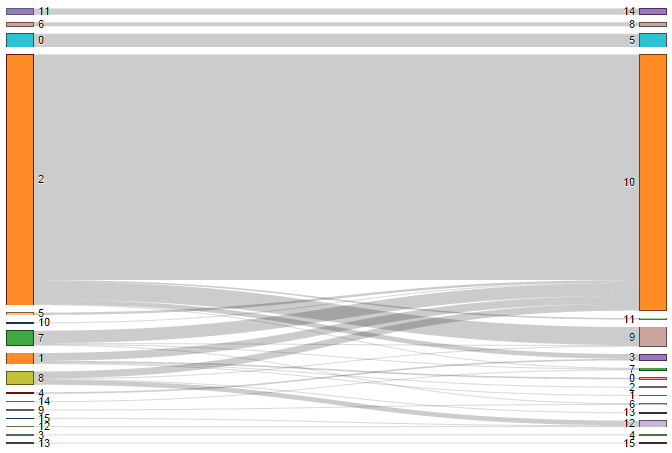}
      \caption*{$D^P$ ARI 0.46}
   \end{minipage} \hfill
   \begin{minipage}[c]{.24\linewidth}
      \includegraphics[width=\textwidth]{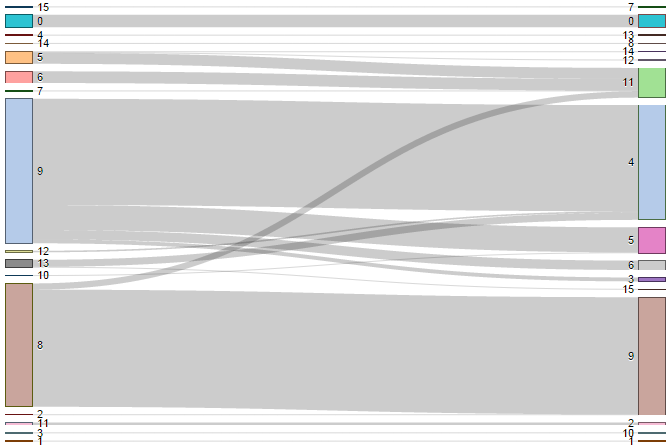}
      \caption*{$D^S$ ARI 0.71}
   \end{minipage} \hfill
   \begin{minipage}[c]{.24\linewidth}
      \includegraphics[width=\textwidth]{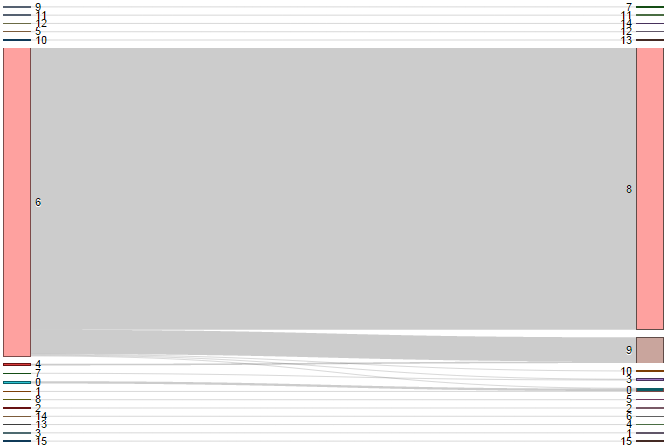}
      \caption*{$D^E$ ARI 0.47}
   \end{minipage} \hfill
   \begin{minipage}[c]{.24\linewidth}
      \includegraphics[width=\textwidth]{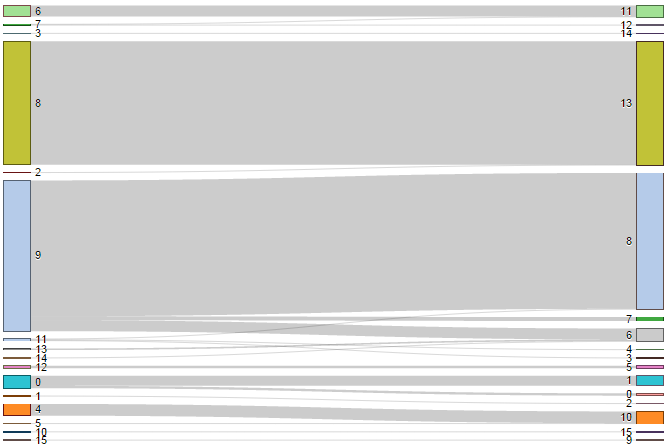}
      \caption*{$D^G$ ARI 0.90}
   \end{minipage}
   \caption{Stability to Odd vs. Even perturbation and the associated ARI showing a better stability of $D^G$-based clustering; partitions obtained from $D^S$-based clustering are rather similar but less stable}\label{fig:oddvseven}
\end{figure}

In Fig. \ref{fig:oddvseven}, we display the stability results on the Odd vs. Even experiments. For each distance, we have displayed the partitions obtained on the odd trading days sample (left) and on the even trading days sample (right). A grey link binds a given asset in the left partition to the same asset in the right partition. Thus, a perfectly stable clustering is displayed by a one-to-one correspondence between left and right clusters. Diverging edges highlight mismatches between partitions, hovering on the edges shows the assets switching from clusters. In this experiment, these can be assets with an unusual history, for instance they  may have encountered a strong variation on a particular day due to a merger (M\&A), a catastrophe or a fraud. But, of course, a cluster switch can happen due the clustering method shortcomings.

\begin{figure}
   \begin{minipage}[c]{.24\linewidth}
      \includegraphics[width=\textwidth]{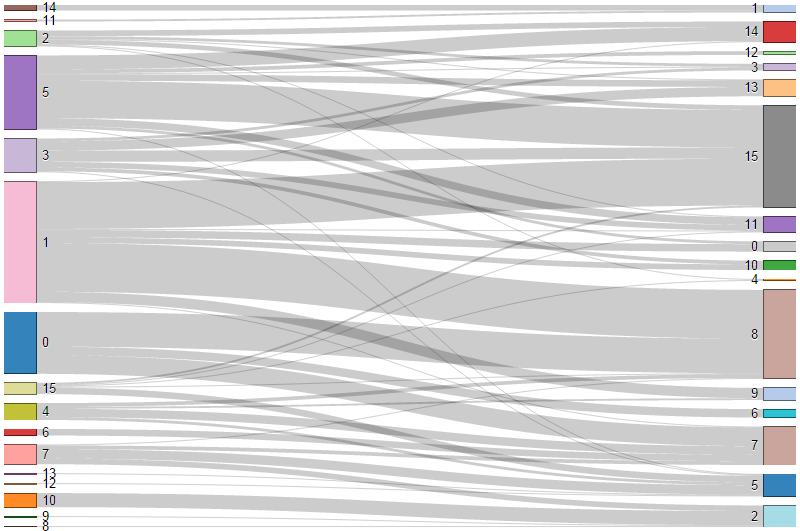}
      \caption*{$D^P$ ARI 0.24}
   \end{minipage} \hfill
   \begin{minipage}[c]{.24\linewidth}
      \includegraphics[width=\textwidth]{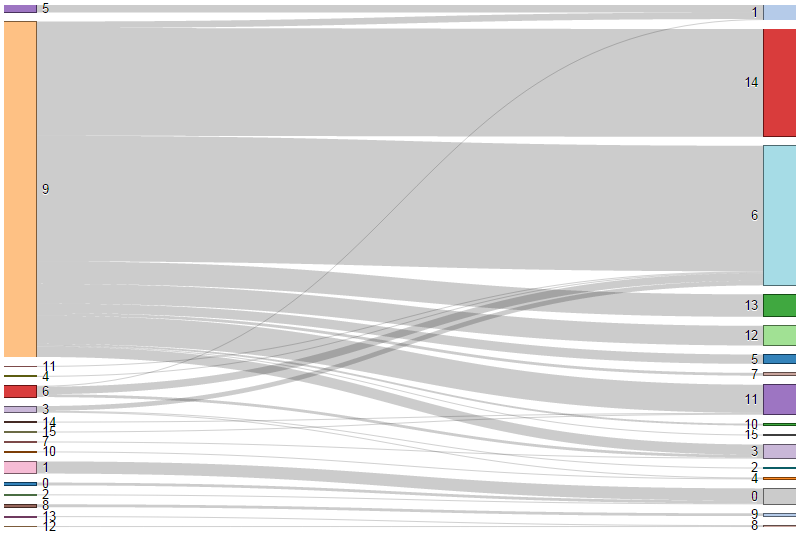}
      \caption*{$D^S$ ARI 0.09}
   \end{minipage} \hfill
   \begin{minipage}[c]{.24\linewidth}
      \includegraphics[width=\textwidth]{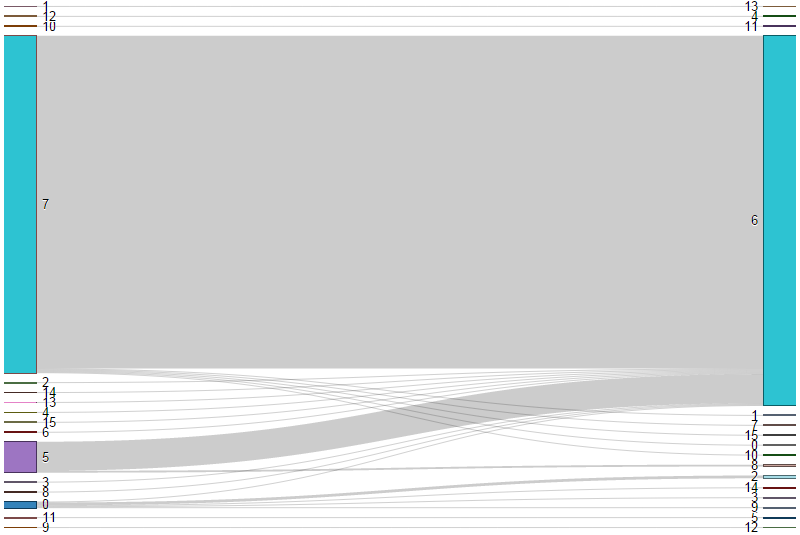}
      \caption*{$D^E$ ARI 0.31}
   \end{minipage} \hfill
   \begin{minipage}[c]{.24\linewidth}
      \includegraphics[width=\textwidth]{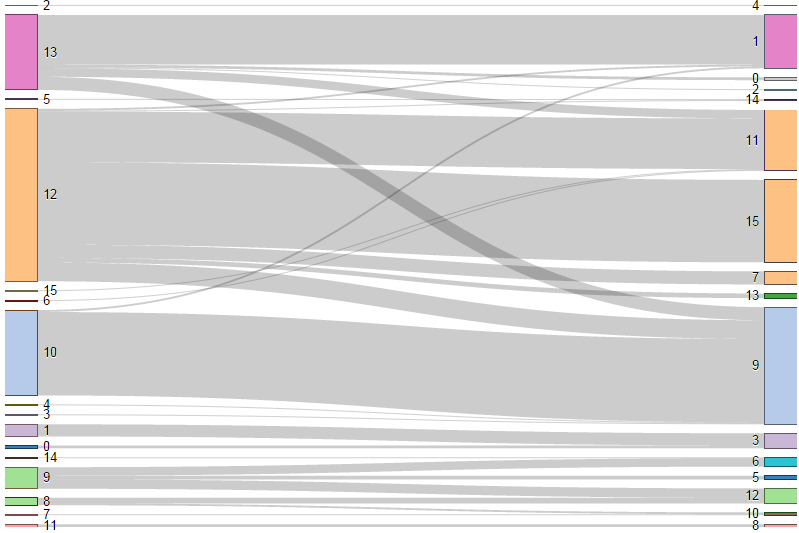}
      \caption*{$D^G$ ARI 0.43}
   \end{minipage}
   \caption{Stability to Heart vs. Tails perturbations for different distances and the associated ARI}\label{fig:heartvstails}
\end{figure}

The Heart vs. Tails experiment displayed in Fig. \ref{fig:heartvstails} shows an interesting stylized fact about the CDS market. Clustering on correlation ($D^S$ and $D^P$) is not stable at all with respect to this perturbation. This means that the sample of the strongest moves in the market has a totally different clustering structure than the sample of the mildest moves when considering only correlation. This can be explained since when the market is stressed, macroeconomic tensions tend to affect all the participants and correlation between assets becomes significatively higher and similar for all assets, thus becomes uninformative. This claim is supported by the fact that $D^S$ based on the Spearman correlation (correlation between ranks) performs the worst, whereas $D^P$, based on the Pearson correlation measure known to be decreased by fat-tailed variations, achieves a better stability since this correlation-based distance discriminates unintentionally on distributions. For high values of $\rho$, which is the case in stressed period, $D^E$ discriminates on the mean and variance of the variations, so performs better than the correlation-based distances. Finally, $D^G$ which intentionally works on both information can leverage the distribution information and obtain a rather stable clustering between the stress periods and the more quiet ones.

\begin{figure}
   \begin{minipage}[c]{.24\linewidth}
      \includegraphics[width=\textwidth]{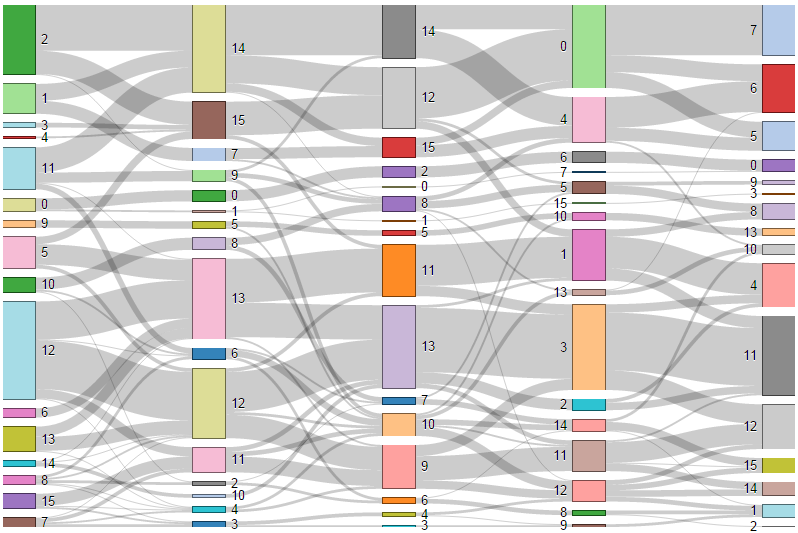}
      \caption*{$D^P$}
   \end{minipage} \hfill
   \begin{minipage}[c]{.24\linewidth}
      \includegraphics[width=\textwidth]{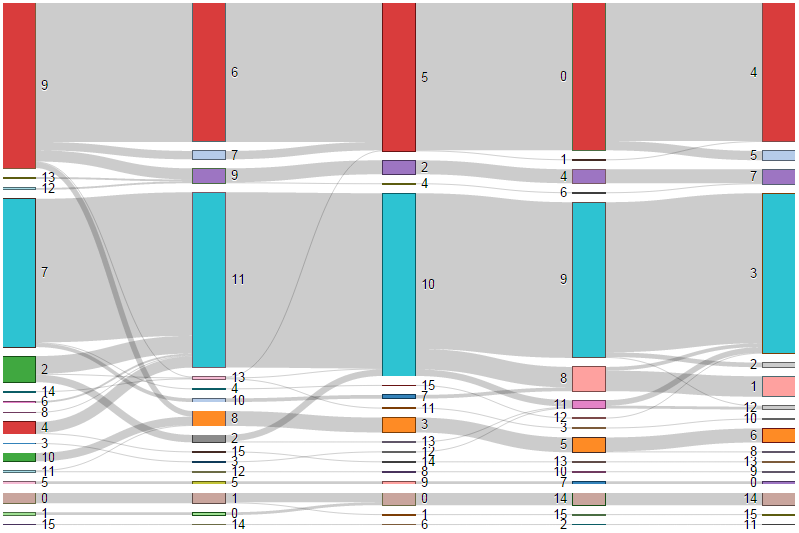}
      \caption*{$D^S$}
   \end{minipage} \hfill
   \begin{minipage}[c]{.24\linewidth}
      \includegraphics[width=\textwidth]{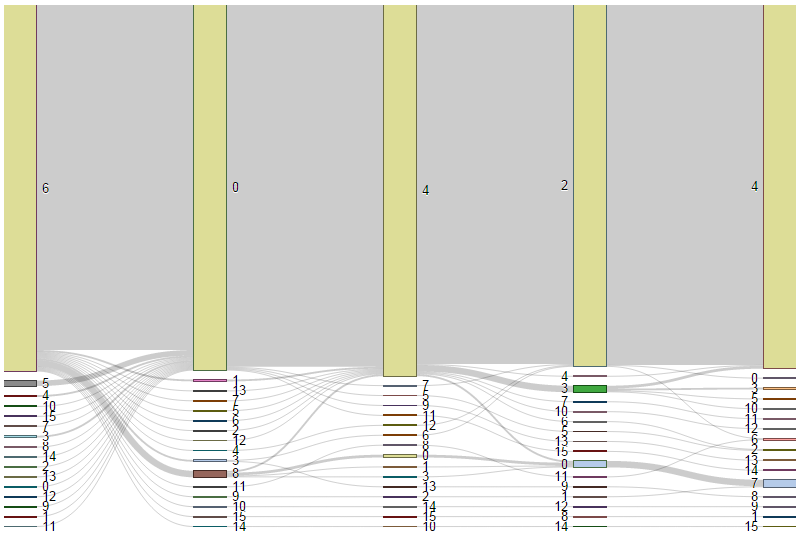}
      \caption*{$D^E$}
   \end{minipage} \hfill
   \begin{minipage}[c]{.24\linewidth}
      \includegraphics[width=\textwidth]{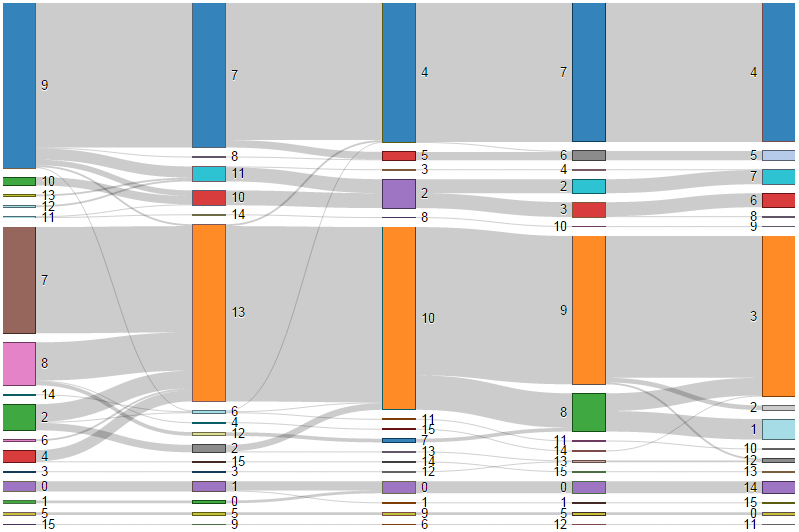}
      \caption*{$D^G$}
   \end{minipage}
   \caption{Stability to Maturity perturbations}\label{fig:maturity}
\end{figure}

In Fig. \ref{fig:maturity}, we display results of the Maturity experiment. For each clustering, we show 5 partitions corresponding to clustering the 1,3,5,7,10-year CDS. We can notice that the partition corresponding to the 1-year CDS is the less stable whatever the distance used. This can be explained by the relative illiquidity of the 1-year maturity compared to the others yielding to scarce and noisy quotes from the market makers. Stability is high for $D^S$ and $D^G$ and abnormally low for $D^P$ and $D^E$ while information is essentially the same.

\begin{figure}
   \begin{minipage}[c]{.24\linewidth}
      \includegraphics[width=\textwidth]{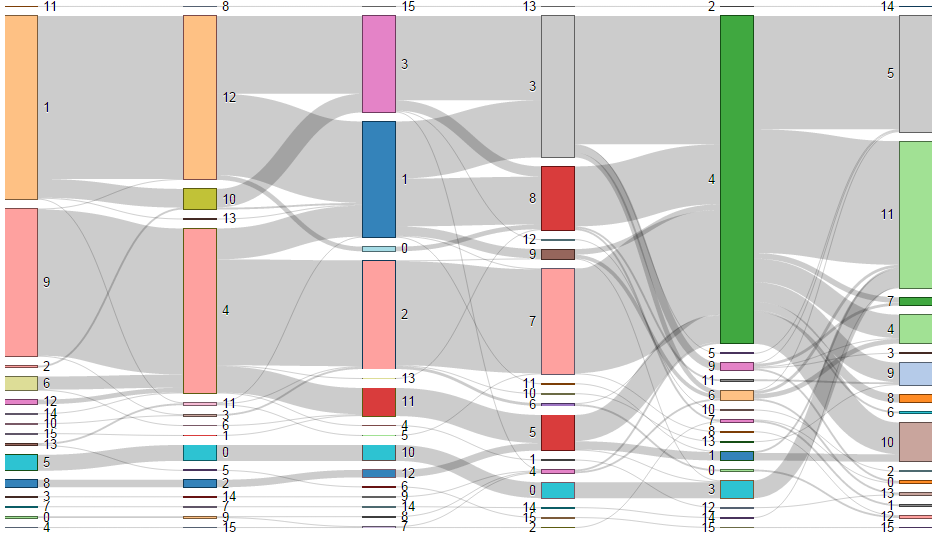}
      \caption*{$D^P$}
   \end{minipage} \hfill
   \begin{minipage}[c]{.24\linewidth}
      \includegraphics[width=\textwidth]{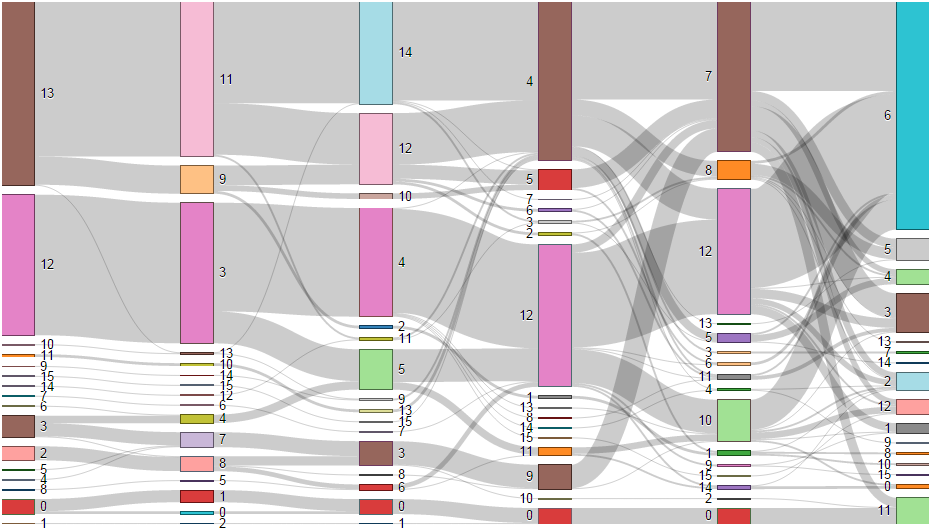}
      \caption*{$D^S$}
   \end{minipage} \hfill
   \begin{minipage}[c]{.24\linewidth}
      \includegraphics[width=\textwidth]{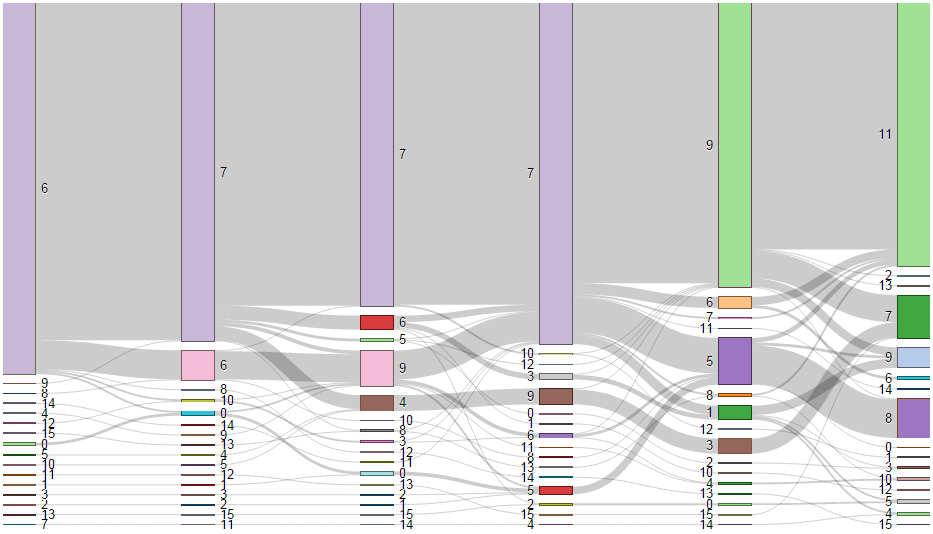}
      \caption*{$D^E$}
   \end{minipage} \hfill
   \begin{minipage}[c]{.24\linewidth}
      \includegraphics[width=\textwidth]{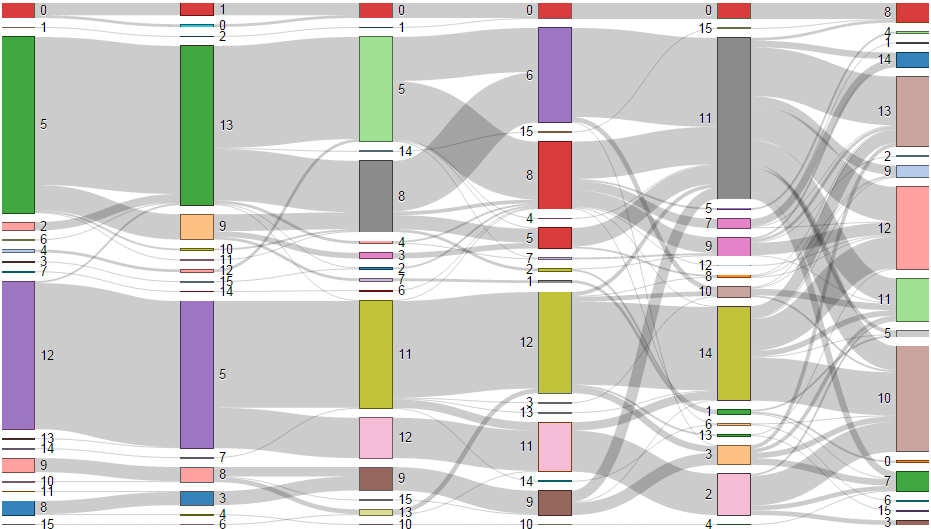}
      \caption*{$D^G$}
   \end{minipage}
   \caption{Stability to Multiscale perturbations}\label{fig:multiscale}
\end{figure}

The Fig. \ref{fig:multiscale} depicts results of the Multiscale experiments. 6 partitions are displayed for each distance corresponding to the clusterings obtained by considering respectively 1,2,4,8,16,32 trading days variations.
We can observe as a stylized fact that the clustering structure is persistent up to a weekly sampling, and that the clustering structure is essentially determined by correlations as advocated by the high stability achieved by $D^S$ and $D^P$. $D^G$, once again, is relatively stable leveraging its correlation part which is similar to $D^S$.

\begin{figure}
   \begin{minipage}[c]{.24\linewidth}
      \includegraphics[width=\textwidth]{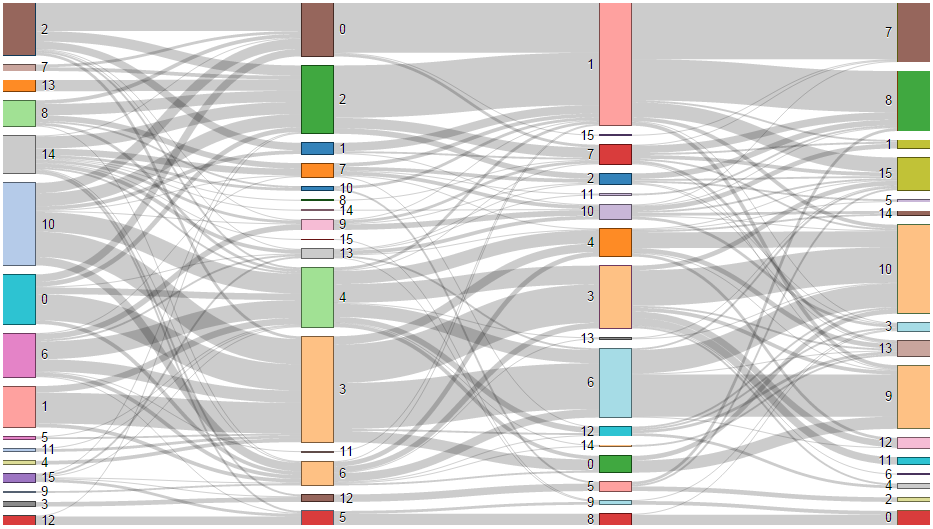}
      \caption*{$D^P$} 
   \end{minipage} \hfill
   \begin{minipage}[c]{.24\linewidth}
      \includegraphics[width=\textwidth]{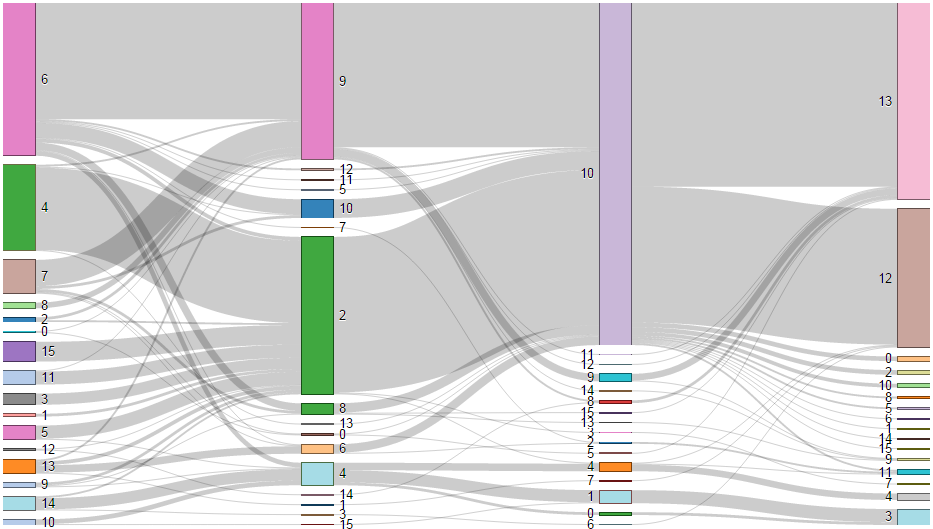}
      \caption*{$D^S$} 
   \end{minipage} \hfill
   \begin{minipage}[c]{.24\linewidth}
      \includegraphics[width=\textwidth]{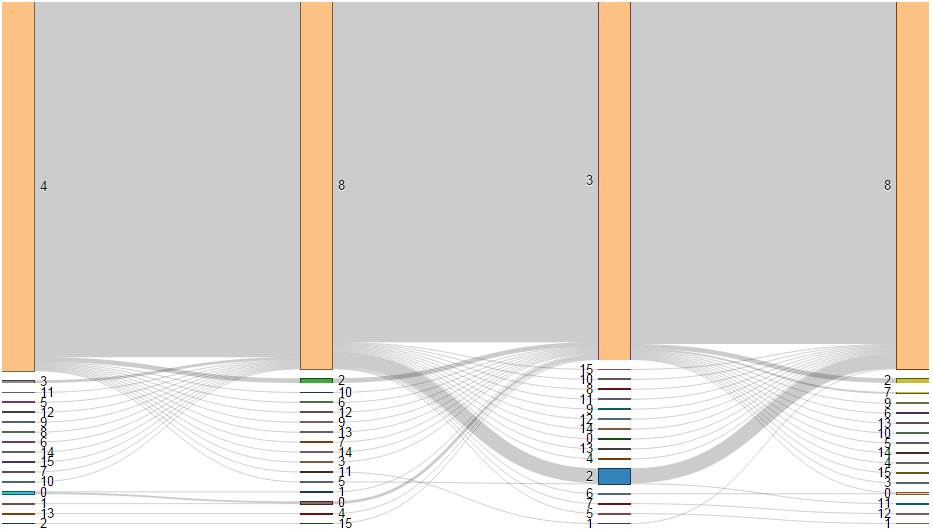}
      \caption*{$D^E$}
   \end{minipage} \hfill
   \begin{minipage}[c]{.24\linewidth}
      \includegraphics[width=\textwidth]{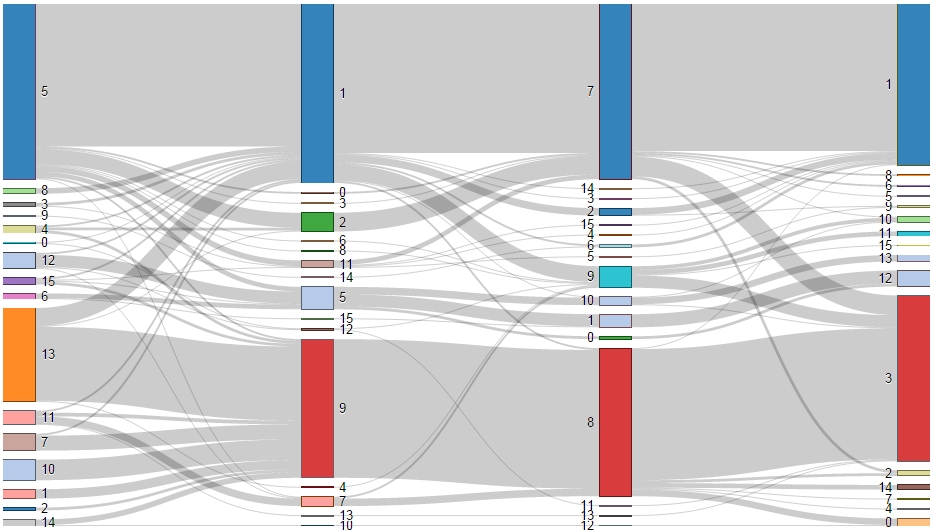}
      \caption*{$D^G$}
   \end{minipage}
   \caption{Stability to Economic Regimes perturbations}\label{fig:ecoreg}
\end{figure}

We finally conclude this empirical study with the Economic Regimes perturbations. In Fig.~\ref{fig:ecoreg}, we display 4 partitions corresponding to the clusterings obtained on different economic periods. From left to right, the pre subprime crisis period 2006-2007, the subprime crisis period 2008-2009, the European debt crisis 2011-2012 and the quantitative easing 2013-2014. We can notice in Fig.~\ref{fig:ecoreg} that the period 2006-2007 yields very different clusters compared to what follows. Indeed, looking at Fig.~\ref{fig:mean_correl}, we observe that correlation in the market was very low. Except clustering with $D^G$, clusters obtained with the other methods are not stable. The partitions and their stability scores obtained from the $D^G$-based clustering agree with previous remarks: pre-crisis period was much different, the clustering structure is the same during both crises, and now that correlation is decreasing and that quantitative easing is at work the clustering structure of the market is changing.

\section{Discussion}
\label{discussion}

In this paper, we have suggested an empirical framework for both assessing clustering validity and investigating the clustering properties of financial time series. This empirical study allows to verify known stylized facts about the financial markets.
Much research work needs to be done before clustering can become a technology in trading rooms or an accepted tool for economic applied research. With this experimental study, we hope to have aroused curiosity from the clustering community, and that further progress will be achieved in the field of clustering financial time series using a framework based on this one for systematically benchmarking the proposed algorithms taking into account the wishes of practitioners.




\bibliographystyle{IEEEtran}
\bibliography{IEEEabrv,IEEEexample}
%

\end{document}